%% file: cospar_proc.tex
          [2001/04/25 1.1 (PWD)]
\documentclass[a4paper,twocolumn]{esapub2005} 
\usepackage{times}
\usepackage{natbib}
\usepackage{amsmath,mathenv,graphicx,color,tensor,amsfonts,bm}

\include{macro}

\title{Mapping the Spacetime Metric with GNSS: a preliminary study}
\author{P. Delva}
\author{J. T. Olympio}
\affil{European Space Agency, The Advanced Concepts Team, Keplerlaan 1, 2201 AZ Noordwijk, The Netherlands}

\begin{document}

\keywords{General relativity; relativistic coordinates; relativistic gravimetry}

\maketitle

\begin{abstract}
In order to do relativistic gravimetry one needs to define a system of null coordinates for a given constellation of satellites. We present here three methods in order to find the null coordinates of an event in a Schwarzschild geometry. We implement these three methods for the weak gravitational field of the Earth, compare their precision and time of computation.
\end{abstract}

\section{Introduction}
The classical concept of positioning system for a Global navigation satellite system (GNSS) would work ideally if all satellites and the receiver were at rest in an inertial reference frame. But at the level of precision needed by a GNSS, one has to consider curvature and relativistic inertial effects of spacetime, which are far from being negligible~\cite{ashby03}. There are at least two very different ways of dealing with relativistic effects.

One way is to try to preserve the Newtonian conception of absolute time and space, by adding corrections coming from general relativity. This leads to numerous corrections that depend on the kinematics of the positioning system (see~\cite{ashby03} for a detailed description). However, the natural evolution of GNSS is to become more and more accurate with the help of very stable clocks, and autonomous (no need for ground stations) with cross links between the satellites. Galileo will embark hydrogen maser clocks with a drift of about 1 ns after one day. But state of the art atomic clocks are far more stable than this (e.g. a drift of 26 ps/day for the Cesium clock Pharao, and only 0.3 ps/day for optical clocks). At this low level of uncertainty a lot of supplementary corrections have to be added~\cite{linet02}.

Another way to define a positioning system is to abandon the Newtonian concept of absolute space and time, which is known to be a classical approximation, and to define a relativistic positioning system, with the so-called emission, GPS or null coordinates~\cite{blagojevic02,coll06a,rovelli02}. A project called SYPOR, "SYstème de POsitionnement Relativiste", has been proposed by B. Coll and collaborators~\cite{coll03a,pascualsanchez07a}. The relativistic positioning system is composed of four satellites broadcasting their proper times by means of electromagnetic signals. The coordinates of the receiver are simply the four proper times $\{ \tau_1,\tau_2,\tau_3,\tau_4 \}$ received from the four satellites. Then the receiver knows his trajectory in these null coordinates. Moreover, if each satellite broadcasts its proper time to the other satellites, the receiver knows also the trajectories of the satellites, and the system is auto-locating.

The null coordinates have very attractive properties. First of all, they are covariant quantities, although dependent on the set of satellites one chooses and on their trajectories~\cite{lachiezeray06}. The set of satellites constitutes a primary reference system, with no need to define a terrestrial reference frame; so there is no need to track the satellites with ground stations or to synchronize the clocks. There is no need also for relativistic corrections, as relativity is already included in the definition of the positioning system. If needed, they can be related to more usual terrestrial coordinate system. Coll and collaborators~\cite{coll06b,coll06c} studied such relativistic positioning systems in the case of a two-dimensional spacetime, for geodesic emitters in a Minkowski plane and for static emitters in the Schwarzschild plane. The relativistic positioning system has been studied in the vicinity of the earth, performing calculations at first order curvature in a Schwarzschild spacetime~\cite{bahder01, ruggiero08}. On one hand, such an approach at first order does not take advantage of the full meaning of a relativistic positioning system, and suppose that an underlying (non physically sounded) coordinate system is predefined; on the other hand, it has the advantage to give a model of the spacetime geometry in the vicinity of the earth, to which the data of a relativistic positioning system can be compared.

The next generation of GNSS will have cross-link capabilities. Each satellite will broadcast the proper time of the other satellites in view, as well as their proper time. With these informations, one could \emph{in principle} map the spacetime geometry in the vicinity of the constellation of satellites by solving an inverse problem~\cite{tarantola09}. This proposal is at the origin of this study. Here, the use of relativistic coordinates is not proposed to enhance the precision of a positioning system, but to achieve a scientific objective, which is the mapping of the gravitional field in the vicinity of a GNSS. In the inverse problem, the basic unknowns are the components of the spacetime metric in the null coordinate system. A primary constellation, consisting of four satellites, defines the null coordinate system; the secondary constellation is constituted by the satellites that do not contribute to define the null coordinates. The data of the problem are mainly the proper times of the satellites broadcasted to the other satellites. These signals are electromagnetic waves, so their trajectories are null geodesics. Knowing the trajectory of the satellites and the proper time of emission of a signal, one can solve the geodesic equation and compute a proper time of arrival; by comparing the computed proper time with the observed one we can optimize our knowledge of the spacetime metric. In case the satellites are not in free-fall, additional data coming from an accelerometer, a gyrometer or a gradiometer could be used. Then the reconstituted geometry can be compared to a parametrized relativistic model of the spacetime geometry around the earth.

As the existing GNSS are not autonomous, the required data to do a mapping of the gravitational field are missing. We thus propose as a project going beyond this article to simulate these data, and to apply an inverse problem in order to recover the initial metric. In the first part of this article we introduce basic results of null coordinates in a flat spacetime. In the second part we present three methods in order to solve numerically the time transfer problem in a Schwarzschild spacetime: the first method relies on the world function~\cite{synge60}, and requires the implementation of a numerical integration algorithm and a shooting method; the second method relies on an analytic solution of the geodesic equations in terms of elliptic functions~\cite{cadez05}; and the third method relies on series in power of $G$ - the gravitational constant - of the time transfer function~\cite{teyssandier08}. In the third part of this paper we apply these three methods to find the null coordinates of a stationary point in a Schwarzschild spacetime, we compare them in terms of precision and time of computation.

\section{Basic results in flat spacetime}
The null coordinates in a four dimensional flat spacetime has been studied by several authors (\cite{bini08,coll09a,blagojevic02,rovelli02,ruggiero08}). We summarize here the main results. Let call $\w{x} \equiv (x^1,x^2,x^3,x^4)$ the usual minkowskian coordinates, where $x^4 = c t$. The spacetime metric in a flat spacetime is the Minkowski metric $\eta_{\alpha \beta} \equiv \text{diag} [ -1 , -1, -1, 1]$ and 
\be \dd s^2 = \eta_{\alpha \beta} \dd x^\alpha \dd x^\beta . \ee
We consider four satellites $A=\{ 1..4 \}$ that constitute a relativistic positioning system. We assume that each satellite has on board a perfect clock that delivers its proper time $\tau^A$. Then the geodesic equation for the trajectory $\w{x_A} (\tau^A)$ of satellite $A$, parametrized by its proper time, is
\be \dfrac{\dd^2 \w{x_A}}{\dd (\tau^A)^2} = 0 . \ee
The solution of which is:
\be \w{x_A} (\tau^A) = \w{U_A} \tau^A + \w{S^0_A} , \ee
where $\w{U_A} = \dd \w{x_A} / \dd \tau^A$ is a normal constant vector\footnote{Normal means that $\eta_{\alpha \beta} U^\alpha_A U^\beta_A = 1$.}. To simplify further we assume that $\w{S^0_A} = \w{0}$.

\begin{figure}[ht] 
\begin{center}
      \includegraphics[width=0.8\linewidth]{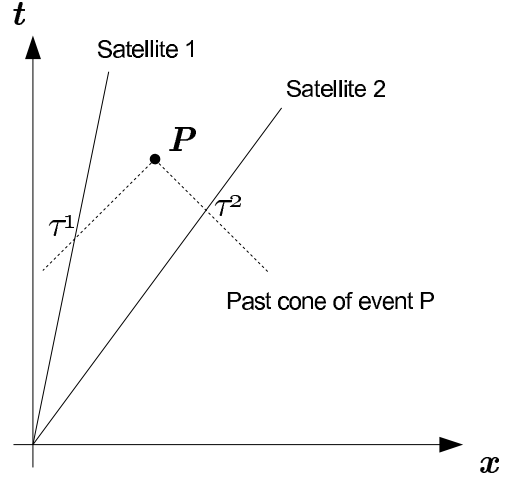}
      \caption{\label{2DFLAT} \footnotesize Illustration of the problem in two dimensions: in order to find the null coordinates of point $P$, one has to find the intersection of its past cone with the satellite trajectories.}
\end{center}
\end{figure}

Let $P$, the point of coordinates $\w{x_P}$, be the receiver location. To find its null coordinates we need to find the null geodesics linking this point to the four satellite trajectories (see Figure~\ref{2DFLAT}). To do this one can use the world function~\cite{synge60} (see appendice~\ref{a:DEF} for a definition). The condition for the signal emitted by the four satellites to meet the receiver location is
\be \label{FLATCOND} \Omega (\w{x_A} (\tau^A), \w{x_P}) = 0 \ , \ x^4_A (\tau^A) < x^4_P . \ee
In flat spacetime the world function is
\be \label{FLATWF} \Omega (\w{x_P},\w{x_A}) = \dfrac{1}{2} \eta_{\alpha \beta} \left( x_P^\alpha - x_A^\alpha \right) \left( x_P^\beta - x_A^\beta \right) . \ee
Then, we find
\be \label{FLATSYS} \vectornorm{x_P}^2 - 2 \tau^A \scalarprod{x_P}{U_A} + \left( \tau^A \right)^2 = 0 , \ee
where we define $\scalarprod{a}{b} \equiv \eta_{\alpha \beta} a^\alpha b^\beta$ and $\vectornorm{a}^2 \equiv \scalarprod{a}{a}$.
We choose the solution of this equation which satisfies the inequality in~\eqref{FLATCOND}. It defines the null coordinates:
\be \tau^A = \scalarprod{x_P}{U_A} - \sqrt{\scalarprod{x_P}{U_A} - \vectornorm{x_P}^2} \ee
This relation is true for any point $P$ of spacetime: then the coordinate transformations $\w{\tau} = \w{\tau} (\w{x})$ (where $\w{\tau} \equiv ( \tau^1, \tau^2, \tau^3, \tau^4 )$) are known everywhere, and the components of the contravariant metric in the null coordinates are
\be g^{A B} = \eta^{\alpha \beta} \dfrac{\partial \tau^A}{\partial x^\alpha} \dfrac{\partial \tau^B}{\partial x^\beta} . \ee

In order to find the components of the covariant metric $g_{A B}$ one can find the inverse metric, or solve the system~\eqref{FLATCOND} with $\w{x_P}$ as the unknown. As Bini \emph{et al.} emphasized~\cite{bini08}, it is equivalent to solve the system:
\begin{eqnarray}
\left( \tau^1 \right)^2 + \vectornorm{x_P}^2 - 2 \tau^1 \scalarprod{x_P}{U_A} & = & 0 \label{e:SYS1} \\
\left( \tau^i \right)^2 - \left( \tau^1 \right)^2 - 2 \left( \tau^i - \tau^1 \right) \scalarprod{x_P}{U_A} &=& 0 \label{e:SYS2}
\end{eqnarray}
where $i=2..4$. In the system of equations \eqref{e:SYS1}-\eqref{e:SYS2} there is only one equation of second order, instead of four for the system~\eqref{FLATCOND}. Then
\be g_{A B} = \eta_{\alpha \beta} \dfrac{\partial x^\alpha}{\partial \tau_A} \dfrac{\partial x^\beta}{\partial \tau_B} . \ee

Alternatively, one can use the time transfer function~\cite{teyssandier08} to find the null coordinates of point $P$ (see appendice~\ref{a:DEF} for a definition). In a flat spacetime, the time delay function is trivial (eq.~\eqref{FLATTT}) so that
\be x^4_P - x^4_A (\tau^A) = R_{AP} (\tau^A) , \ee
which is equivalent to the equations~\eqref{FLATSYS}.

\section{Solving the time transfer in a Schwarzschild geometry}

The spacetime metric written in the usual Schwarzschild coordinates $\w{x} \equiv (r,\theta,\phi,ct)$ is
\bea \dd s^2 &=& g_{\alpha \beta} \dd x^\alpha \dd x^\beta  \\
&=& A(r) \dd t^2 - A^{-1}(r) \dd r^2 - r^2 \dd \Omega^2 \eea
where
\be A(r) = \left( 1-\dfrac{r_S}{r} \right) , \ee
$\dd \Omega^2 = \left( \dd \theta^2 + \sin^2 (\theta) \dd \phi^2 \right)$ and $r_S$ is the Schwarzschild radius.

Let call $\w{x_P}$ the coordinates of the point of reception $P$, and $\w{S_A} = \w{x_A} (\tau^A)$ the worldline of satellite A, parametrized by its proper time $\tau^A$. 

\subsection{Method 1}
The first method to find the null coordinates $\w{\tau}$ of point $P$ uses the world function: one has to solve the system of equations~\eqref{FLATCOND}. This system here is equivalent to
%
\be \int_0^1 \dd \lambda \left[ g_{\alpha \beta} \left( \w{x_L}(\lambda) \right) \dot{x}^\alpha_L (\lambda) \dot{x}^\beta_L (\lambda) \right] = 0 , \ee
where $\lambda$ is an affine parameter, chosen such that $\w{x_L} (0) = \w{x_A} (\tau^A)$ and $\w{x_L} (1) = \w{x_P}$, $\dot{()}\equiv \dd / \dd \lambda$, and $\w{x_L} (\lambda)$ is solution of the geodesic equations:
%
%
\be \dfrac{\nabla \w{u_L}}{\dd \lambda} = 0 ,\ee
%
where $\nabla$ is the covariant derivative and $\w{u_L} = \dd \w{x_L} / \dd \lambda$. This forms a two point boundary value problem that can be solved using a shooting algorithm (see~\cite{sanmiguel07}).

\subsection{Method 2}
The second method is to solve the time transfer equation using the analytic solution of the geodesic equations in terms of elliptic functions~\cite{cadez05}. We summarize here the main results. The differential equation for orbits of lightlike geodesics is
\be \dfrac{\dd u}{\dd \lambda} = \pm \sqrt{a^2 - u^2 (1-u)} , \ee
where $u = r_S / r$, $\lambda$ is the true anomaly of the orbit and $a$ is a constant of motion. The solution of this equation for type A orbits (ie. scattering orbit with both end points at infinity) can be expressed in terms of elliptic functions with the transformation $u = t^2 (u_2-u_3)+u_3$:
\be \label{e:ELL} u = u_2 - (u_2-u_3) \text{cn}^2 \left[ \left( F(\chi_A | m) + \dfrac{\lambda - \lambda_A}{n} \right) | m \right] , \ee
where $m=(u_2-u_3)/(u_1-u_3)$, $n=2/\sqrt{u_1-u_3}$, and $u_1$, $u_2$ and $u_3$ are solution of the cubic equation $a^2-x^2(1-x)=0$; $\lambda_A$ is the true anomaly of a point $A$ lying on the trajectory, and $\chi_A$ is such that
\be \sin^2 \chi_A = \dfrac{u_A-u_3}{u_2-u_3} ,\ee
where $u_A=r_S / r_A$, with $r_A$ the radial coordinate of point $A$; cn is the usual elliptic function and $F$ is the incomplete elliptic integral of the first kind.

If the spatial coordinates of two different points $A$ and $P$ are given, then the parameter $a$ of the null geodesic crossing these two points can be found by solving a nonlinear equation on $a$, which can be found by using the Jacobi elliptic functions addition theorem and the general expression~\eqref{e:ELL} of the orbit (eq.(19) of~\cite{cadez05}). Once the parameter $a$ of the null geodesic between the points $A$ and $P$ is known, one can calculate the time of flight $T_f (\vx_P,\vx_A;a)$ of the photon between $A$ and $P$ by using the formula~(25) of~\cite{cadez05}. If the arrival coordinate time $x^4_P$ is given, and that the emission point $A$ lies on the given orbit $\w{x_A} (\tau^A)$ of satellite $A$, then one can find the null coordinates $\w{\tau}$ of $P$ by solving numerically
\be x^4_P - x^4_A (\tau^A) = T_f (\vx_P,\vx_A (\tau^A);a) . \ee

\subsection{Method 3}
The third and last method uses a post-Minkowskian expansion of the time transfer function~\cite{teyssandier08} for a static and spherical body. The second order post-Minkowskian expansion of the time delay function between two points $A$ and $P$ (see appendice~\ref{a:DEF}) is then (for general relativity):
\bea
\Delta (\vx'_A, \vx'_P) &=& r_S \ln \left( \dfrac{r'_A + r'_P + R'_{AP}}{r'_A + r'_P - R'_{AP}} \right) \nonumber \\ 
&+& r_S^2 \dfrac{R'_{AP}}{r'_A r'_P} \left[ \dfrac{15}{4} \dfrac{\arccos (\vn'_A \cdot \vn'_P)}{\sqrt{1-(\vn'_A \cdot \vn'_P)^2}} \right. \nonumber \\
&-&  \left. \dfrac{4}{1+(\vn'_A \cdot \vn'_P)} \right] \label{e:TT2} \\
&+& \Ol(r_S^3) ,
\eea
where $\w{x}'$ are the quasi-Cartesian isotropic coordinates, $r'_A = \vectornormeucl{\vec{x}'_A}$ (with the euclidean norm), $r'_P = \vectornormeucl{\vec{x}'_P}$, $R'_{AP} = \vectornormeucl{\vx'_A - \vx'_P}$, $\vn'_A=\vx'_A / r'_A$, $\vn'_P=\vx'_P / r'_P$ and $r_S$ is the Schwarzschild radius. This function is independant of the time of reception and the time of emission, as it should be for a stationary spacetime.

The transformation from the Schwarzschild coordinates to the quasi-Cartesian isotropic coordinates can be obtained with: $t'=t$, $\theta'=\theta$, $\phi'=\phi$ and (up to the second order in $(r_S / r)$)
\be r' = r \left[ 1 - \dfrac{r_S}{2 r} - \left( \dfrac{r_S}{4 r} \right)^2 + \Ol \left( \dfrac{r_S}{r} \right)^3 \right] . \ee
Then, for a given point P of coordinates $\w{x_P}$, one obtains its null coordinates $\w{\tau}$ by resolving the four equations:
\be x^4_P - x^4(\tau^A) = R_{AP} + \Delta \left( \vx (\tau^A), \vx_P \right) , \ee
where $\w{x}(\tau^A)$ is the given orbit of satellite $A$.

Note that this method is only valid to second order in $(r_S/r)$. It is possible to obtain the higher order terms in equation~\eqref{e:TT2} but at the price of complicated calculations. On the contrary, the precision of the two first method is only limited by the time of computation of the numerical method used to solve the problem. In the next part we compare the numerical implementation of these three methods to find the null coordinates of a stationary point in a Schwarzschild spacetime.

\section{The null coordinates of a stationary point in Schwarzschild spacetime}
\subsection{The constellation of satellites}
We want to study a constellation of four satellites in the plane $\theta=\pi /2$, with circular orbits, and that define a null coordinate system. The geodesic equations for the satellites are~\cite{papapetrou74}:
\bea
r&=&r_0\\
\ddot{t}&=&0\\
\ddot{\phi}&=&0\\
\dot{\phi}^2&=&\dfrac{r_S}{2 r_0^3} c^2 \dot{t}^2
\eea
where $\dot{()} \equiv \dd / \dd \tau$, with $\tau$ an affine parameter. Let $\tau$ be the proper time, then we have the supplementary equation $u_\alpha u^\alpha = 1$, where $u^\alpha=\dd x^\alpha / \dd \tau$. The solution is:
\bea
r&=&r_0 \label{e:TRAJSAT1}\\
t (\tau)&=&\dfrac{1}{\sqrt{1-\frac{3 r_S}{2 r_0}}} \tau + t_0 \label{e:TRAJSAT2}\\
\phi (\tau)&=&\sqrt{\dfrac{c^2 r_S}{2 r_0^3 \left( 1-\frac{3 r_S}{2 r_0} \right)}} \tau + \phi_0 \label{e:TRAJSAT3}
\eea
where $t_0$ and $\phi_0$ are two constant of integration. No circular orbits exist for $r_0\le 3 r_S / 2$. It is also well-known~\cite{misner73} that for $3 r_S / 2 < r_0 \le 3 r_S$, circular orbits are unstable, and stable for $r_0 > 3 r_S$.
One can notice that
\be \dfrac{\dd \phi}{\dd t}=\sqrt{\dfrac{c^2 r_S}{2 r_0^3}} \ee

In order to compare the different methods sketched in the previous section, we need to solve the problem for only one satellite $A$ of the constellation. Its initial conditions are $r_0$, $t_0$ and $\phi_0$. To simplify, we choose $t_0=0$ and $\phi_0=0$, and we note $r_0$ and $\tau^A$ the radius and the parameter of the satellite trajectory. Let $P$ be a point of given coordinates $\w{x_P}$, and $\w{x_A}(\tau^A)$ the trajectory of the satellite.


\input{numerical.tex}

\section{Conclusion}
In this paper we recalled the basic properties of null coordinates in flat spacetime. We presented three methods in order to find the null coordinates of an event in a Schwarzschild spacetime. We implemented these three methods and compared them in terms of time of computation and of precision. The post-Minkowskian expansion (method 3) is faster than a direct integration (method 1) or than an evaluation of the elliptic functions (method 2). However using only double precision in the calculation leads to time of calculation that are of the same order. We still need to implement the three methods in multi-precision to compare them to the maximum accuracy of the third method. To go beyond this preliminary study, we want to use these results in order to simulate the data of a constellation of satellites in the context of relativistic gravimetry.

\appendix

\section{Definitions}
\label{a:DEF}
\paragraph{The world function~\cite{synge60}} Let $P$ and $Q$ two points of coordinates $\w{x_P}$ and $\w{x_Q}$, joined by a geodesic $\Gamma : \w{x} = \w{\xi} (\lambda)$ where $\lambda$ is an affine parameter; then the world function $\Omega$ is
\be \Omega (\w{x_P},\w{x_Q}) = \dfrac{1}{2} \int_0^1 g_{\alpha \beta} (\w{x}) |_\Gamma U^\alpha U^\beta \dd \lambda , \ee
where $\w{U} = \dd \w{\xi} / \dd \lambda$ and $\lambda$ is such that $\w{\xi} (0) = \w{x_P}$ and $\w{\xi} (1) = \w{x_Q}$.

\paragraph{The time transfer function~\cite{teyssandier08}} We assume that spacetime is globally regular and without horizon, and that the coordinate system is chosen such that $g_{44} > 0$ everywhere. The past null cone at a given point $P$ of coordinates $\w{x_P}$ intersects the worldline $S_A$ at one and only one point $\w{x_A} = (\vx_A, c t_A)$ (where $\vx_A \equiv (x_A^1,x_A^2,x_A^3)$). The difference $t_P - t_A$ is the (coordinate) travel time of a light ray connecting the emission point $\w{x_A}$ and the reception point $\w{x_P}$. This quantity may be considered either as a function of the instant of reception $t_P$ and of $\vx_A$, $\vx_P$, or as a function of the instant of emission $t_A$ and of $\vx_A$ and $\vx_P$. So it is possible to define two time transfer functions, $\T_r$ and $\T_e$ with:
\be t_P - t_A = \T_r (\vx_A, t_P, \vx_P) = \T_e(t_A, \vx_A, \vx_P) . \ee
$\T_r$ is the reception time transfer function and $\T_e$ the emission time transfer function. These functions are distinct except in a stationary spacetime in which the coordinate system is chosen so that the metric does not depend on $x^4$. In this case the subscript r and e can be omitted.

We suppose now that the metric takes the form
\be g_{\alpha \beta} = \eta_{\alpha \beta} + h_{\alpha \beta} . \ee
Then the reception time transfer function may be written as
\be \T_r (\vx_A, t_P, \vx_P) = \dfrac{1}{c} R_{AP} + \dfrac{1}{c} \Delta_r (\vx_A, t_P, \vx_P) , \ee
where $R_{AP} = \vectornormeucl{\vx_A - \vx_P}$ (with the euclidean norm). The function $\Delta_r / c$ is called the reception time delay function.

In flat spacetime it is obvious that
\be \label{FLATTT} \Delta (\vx_A, \vx_P) = 0 . \ee

\bibliographystyle{aa}
\bibliography{biblio_rps}

\end{document}

%% file: macro.tex
\newcommand{\be}{\begin{equation}}
\newcommand{\ee}{\end{equation}}
\newcommand{\bea}{\begin{eqnarray}}
\newcommand{\eea}{\end{eqnarray}}

\newcommand{\w}[1]{\bm{#1}}
\newcommand{\vectornorm}[1]{\left|\w{#1}\right|}
\newcommand{\scalarprod}[2]{\left<\w{#1},\w{#2}\right>}
\newcommand{\vectornormeucl}[1]{\left|\left|#1\right|\right|}

	\newcommand{\dd}{\text{d}} 

\newcommand{\T}{\mathcal{T}}

\newcommand{\Ol}{\mathcal{O}}

\newcommand{\vx}{\vec{x}}

\newcommand{\vn}{\vec{n}} 



%% file: numerical.tex
\subsection{Numerical results}

The first algorithm uses a shooting method to solve the Two Point Boundary Value Problem (TPBVP). The TPBVP is formulated as an initial value problem (IVP), where the initial point has to be guessed for the terminal point to satisfy the terminal conditions. The IVP can then be solved using a Newton-Raphson algorithm.
To compute the Jacobian we use finite differences. An alternative can be computing the transition matrix, which would result in the integration of a $9^2$ differential system. When using finite differences, the time step has to be selected carefully. One is always tempted to take the smallest  step possible, although that does not imply a better approximation of the derivatives. One elegant approach is to have an accurate numerical value of the derivatives using the theory on holomorphic functions. Although, the functions have to be extended to the complex space, which is not always readily feasible. Another option is to use automatic differentiation to help to reduce the high burden of calculating the analytical derivatives.\par 
Methods 2 solves two algebraic equations involving Jacobi elliptic functions and elliptic integrals. The equations are unfortunately not well suited numerically to be solved readily with Newton-type solvers. We use the Brent method~\cite{brent73}, which is a secant method algorithm, to solve the problem. Another option would be to expand the integrand to calculate the time of flight $T_f$ (formula (23) of \cite{cadez05}) into a convergent series of analytically integrable functions, which should be more suitable and faster for weak gravitational fields~\cite{cadez05}.\par
Method 3 does not require any integration either. We simply need to solve one algebraic equation that happens to be a polynomial equation of second order. This equation can be solved efficiently with a Newton-type algorithm, where the Jacobian can be computed accurately by finite differences.\par
\vspace{\baselineskip}
In this study we used only double-precision calculation. To have the most accurate results and have a fair comparison between the methods, we should use multi-precision calculation for the three methods, but its implementation is not yet done for the first two methods. In a multi-precision implementation with a 128 bits description of the real number, we should expect at least 30 significant digits.
\par
We computed with the 3 methods the null coordinate $\tau^A$ of the point $P$ of spatial Schwarzschild coordinates 
\begin{align*}
	r_P &= 50000 \ \text{km}\\
	\theta_P &= \pi/2 \\
	\phi_P &= 0 ,
\end{align*}
for different arrival times $t_P$. The initial conditions of satellite $A$ are
\begin{align*}
	r_0 &= 42000 \ \text{km}\\
	t_0 &= 0 \\
	\phi_0 &= 0 ;
\end{align*}
its trajectory is given by equations~\eqref{e:TRAJSAT1}-\eqref{e:TRAJSAT3} and is sampled with about 100 points. We choose the Schwarzschild radius of the Earth.

\begin{figure}[ht]
	\centering
	\includegraphics[width=8cm]{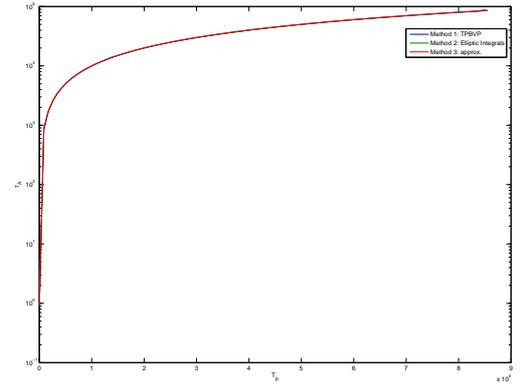}
	\caption{Coordinate $\tau^A$ of the stationary point $P$ as a function of the arrival time $t_P$.}
	\label{fig:tau}
\end{figure}

We use as initial guess, for all three methods, the solution to the flat space problem. This reduces the computation time and ensure a good convergence. The results are shown on figure~\ref{fig:tau}. As the 3 methods provide similar results to the first digits, the graphs are superimposed and no difference is visible.\par 
To compare the methods, we take 4 points sampling the satellite orbits, compute the discrepancies between the solutions for each method, and evaluate the computation time. We take as reference value, the value returned by the method 1.
The results are displayed on table \ref{tab:simulations}.

\begin{table}
	\begin{tabular}{c|c|c||c|c}
				&				&	$\tau^A$ (ref) & & \\
			Pt	& $t_P(s)$	&	Method 1 & Method 1/2 & Method 1/3\\
	\hline
	\hline
		1  &	$10^0$		&$.9733148699$	&	$9.886 1e^{-12}$	&	$7.80 1e^{-15}$\\
	\hline
		2  &	$10^1$	&$9.9733146365$	&	$2.214 1e^{-8}$	&	$1.018 1e^{-13}$\\
	\hline
		3  &	$10^2$	&$99.9732913262$	&  $2.352 1e^{-7}$		&	$8.995 1e^{-12}$\\
	\hline
		4  &	$10^3$	&$999.9710561425$	&	$2.259 1e^{-6}$	&	$7.129 1e^{-11}$\\
	\hline
	\hline
	TC		&			&			1			&			1.25				&			0.5\\
	\end{tabular}
	\caption{Numerical comparisons between the methods}
	\label{tab:simulations}
\end{table}

The computational times (TC) figured on table~\ref{tab:simulations} as ratio with respect to the computational time of method 1. These durations are obtained for double-precision numbers, so not for the most accurate comparison we could do between the different methods.
Assuming that an efficient multi-precision implementation of the methods would lead to similar results, we can infer that method 3 is the fastest. However this method will give results precise up to the second order in $(r_S/r)$, and we are interested in a method that gives a result to any desired accuracy. Methods 1 and 2 are equivalent regarding the precision, but method 1 is hardly faster than method 2. This is mainly because using a Brent algorithm is far from being efficient regarding computation time.

%